\documentclass{article}
\usepackage{spconf,amsmath,graphicx,amssymb,amsfonts}
\usepackage{multirow}
\usepackage{footnote}
\DeclareMathOperator*{\argmax}{argmax}

\title{Enhancing Code-switching Speech Recognition with Interactive Language Biases}
\name{Hexin Liu$^1$, Leibny Paola Garcia$^2$, Xiangyu Zhang$^3$, Andy~W.~H.~Khong$^1$, Sanjeev Khudanpur$^2$}
\address{$^1$School of Electrical and Electronic Engineering, Nanyang Technological University, Singapore\\
  $^2$CLSP and HLT-COE, Johns Hopkins University, USA\\
  $^3$University of New South Wales, Australia}

\begin{document}

\maketitle

\begin{abstract}
 Languages usually switch within a multilingual speech signal, especially in a bilingual society. This phenomenon is referred to as code-switching~(CS), making automatic speech recognition~(ASR) challenging under a multilingual scenario. We propose to improve CS-ASR by biasing the hybrid CTC/attention ASR model with multi-level language information comprising frame- and token-level language posteriors. The interaction between various resolutions of language biases is subsequently explored in this work. We conducted experiments on datasets from the ASRU 2019 code-switching challenge. Compared to the baseline, the proposed interactive language biases~(ILB) method achieves higher performance and ablation studies highlight the effects of different language biases and their interactions. In addition, the results presented indicate that language bias implicitly enhances internal language modeling, leading to performance degradation after employing an external language model.
\end{abstract}

\vspace{-0.2cm}
\section{Introduction}
\label{sec:intro}
Code-switching~(CS) refers to the switching of languages within a spontaneous multilingual recording. Automatic speech recognition~(ASR) faces challenges in a code-switching scenario due to the inter- and intra-sentence language varieties compared to its monolingual counterparts~\cite{kaldi, conformer}. Although conventional ASR approaches can operate on code-switching speech similar to monolingual data, early works identify languages before speech recognition or performs these processes jointly~\cite{firstcsasr,zeng19_interspeech,liuxsa}. In contrast, recent CS-ASR techniques tackle language confusion by incorporating language information in modules within the ASR model. 

One such approach involves the use of a bi-encoder model that is built on the transformer architecture~\cite{transformer, bi_encoder}, where modeling of English and Mandarin languages is decoupled by two encoders pre-trained independently on each language. Since the dual-encoder approach has shown to be language discriminative, CS-ASR approaches that adopted similar architectures were subsequently proposed \cite{mary20_icassp, tt_cs, song22e_interspeech}. Apart from dual encoders, a language-specific attention mechanism has also been proposed to reduce confusion caused by code-switching contexts~\cite{speech_transformer, zhang22x_interspeech}. This attention mechanism is employed within the transformer decoders and processes monolingual token embeddings which are separated from code-switching token sequences. In addition, a conditional factorization method factorizes CS-ASR into two monolingual recognitions before composing recognized monolingual segments into a single bilingual sequence which may or may not be code-switched ~\cite{conditionalfactorization}.

Although existing approaches mitigate the language confusion for CS-ASR, they are generally stuck in only one module within a CS-ASR model. Since language-aware modules have shown to be effective, it is natural to consider incorporating language information in all modules to further enhance the performance of existing approaches. In addition, these approaches utilize language information either at frame-level (dual-encoder methods) or token-level (transformer-decoder-based approaches)~\cite{zhang22x_interspeech,reduce_liu}. Since the ASR process aims to align acoustic frames to texts~(e.g., characters, words), it is desirable to associate frame- and token-level language information and utilize them jointly for CS-ASR.

Inspired by the success of incorporating language information~\cite{reduce_liu}, we propose to enhance language-aware CS-ASR using interactive language biases~(ILB). In particular, the proposed method comprises two contributions. Firstly, we bias the connectionist temporal classification~(CTC), encoder, and decoder modules jointly within a hybrid CTC/attention CS-ASR model with language posteriors. It is useful to note that the language information transits from frames to tokens~(i.e., from the encoder to CTC and decoder) intrinsically. As opposed to existing models, our method utilizes the interaction between frame- and token-level language information resulting in an integrated and language-discriminative model. In addition, the proposed architecture allows the research community to gain insight into how language biases influence a CS-ASR model beyond improving performance. Experiment results suggest that the CS-ASR model is capable of developing a robust internal language model after learning from language information.

\vspace{-0.2cm}
\section{Methodology}
\subsection{Language posterior bias}
\label{sec:relate}
The language posterior bias approach~\cite{reduce_liu} has been developed on the hybrid CTC/attention ASR model, which comprises an encoder module, a decoder module, and a CTC module \cite{hybrid_ctc_attention_asr,ctc}. These encoder and decoder modules consist of conformer encoder layers and transformer decoder layers \cite{conformer, transformer}, respectively. 

Consider a speech signal with its acoustic features $\mathbf{X}=(\mathbf{x}_{t} \in \mathbb{R}^{F}| t=1, \ldots , T)$ and token sequence $W=(w_n \in \mathcal{V} | n=1, \ldots , N)$, where $\mathcal{V}$ is a vocabulary of size $V$, $T$ and $N$ are the lengths of acoustic features and token sequence, respectively. The encoder generates output $\mathbf{H}=(\mathbf{h}_{t} \in \mathbb{R}^{D}| t=1, \ldots , T_{1})$ from $\mathbf{X}$, which are subsequently fed into the decoder and CTC modules. On the other hand, tokens are first embedded into $\mathbf{W}=(\mathbf{w}_{n} \in \mathbb{R}^{D} | n=1, \ldots , N)$ before being fed into the decoder module along with $\mathbf{H}$. 
The ASR model is optimized with a language diarization~(LD) decoder jointly, where the LD decoder computes a $V^{\textrm{ld}}$-dimensional token-level language posterior bias $\mathbf{p}(l_{n-1}|w_{1:n-1},\mathbf{X})$. Here, $V^{\textrm{ld}}$ is the language vocabulary size and $l_{n-1}$ is the language index for the $n$-th token. The token embedding $\mathbf{w}_{n-1}$ is then biased by its language posterior. The ASR decoder output is subsequently computed via
\begin{eqnarray}
  &&\mathbf{H} = \mathrm{Encoder}\left ( \mathbf{X} \right ),
  \label{eq:encoder} 
  \\
  &&\mathbf{w}^{\prime}_{n-1} = \mathrm{Concat}\big (\mathbf{w}_{n-1}, \mathbf{p}\left(l_{n-1}|w_{1:n-1},\mathbf{X}\right) \big ),
  \label{eq:w_prime} 
  \\
  &&p\left(w_{n}|w_{1:n-1},\mathbf{X}\right) = \mathrm{Decoder}\left ( \mathbf{w}^{\prime}_{1:n-1}, \mathbf{H} \right ),
  \label{eq:asr_decoder}
\end{eqnarray}
where $\mathrm{Concat}(\cdot)$ denotes the concatenation operation. The matrix $\mathbf{W}^{\prime}=(\mathbf{w}^{\prime}_{n} \in \mathbb{R}^{D+V^{\mathrm{ld}}} | n=1, \ldots , N)$ consists of input token embeddings of the ASR decoder which are subsequently projected back to $D$ dimensions by a linear layer. The model is optimized via
\begin{equation}
  \mathcal{L}_{\mathrm{joint}}=\alpha \mathcal{L}_{\mathrm{ctc}} + \left ( 1-\alpha  \right ) \mathcal{L}_{\mathrm{att}} + \beta \mathcal{L}_{\mathrm{ld}}
  \label{eq:loss_joint}
\end{equation}
and that the decoding process is similar to (\ref{eq:asr_decoder}) but with input token embeddings $\mathbf{W}$ of the ASR decoder being replaced by $\mathbf{W}^{\prime}$. The integrated model is optimized via a multi-task objective function. Here, $\beta$ is a multi-task learning parameter and $\mathcal{L}_{{\mathrm{ld}}}$ is a label-smoothed cross-entropy loss between the predicted and ground-truth language labels for the LD decoder.
\vspace{-0.2cm}
\subsection{Interactive language biases}
\label{sec:ilb}
We propose to extend the language posterior bias method to frame-level language information. We note that frame-level language identification~(LID) is undesirable since the LID performance generally degrades with shorter speech~\cite{liu22e_interspeech, lre17_perform, tseng2021mandarin}. However, since acoustic frames are tightly associated with tokens in ASR, frame-level language identification over $\mathbf{H}$ may benefit from token-level language diarization in (\ref{eq:w_prime}). The frame-level language posteriors, therefore, enhance the hidden output $\mathbf{H}$ to achieve high language discrimination before being fed into the ASR and LD decoders. Consequently, frame- and token-level language posteriors interact and jointly improve the model performance in CS-ASR.

\begin{figure}[t]
\setlength{\belowcaptionskip}{-1cm}
  \centering
  \includegraphics[width=\linewidth]{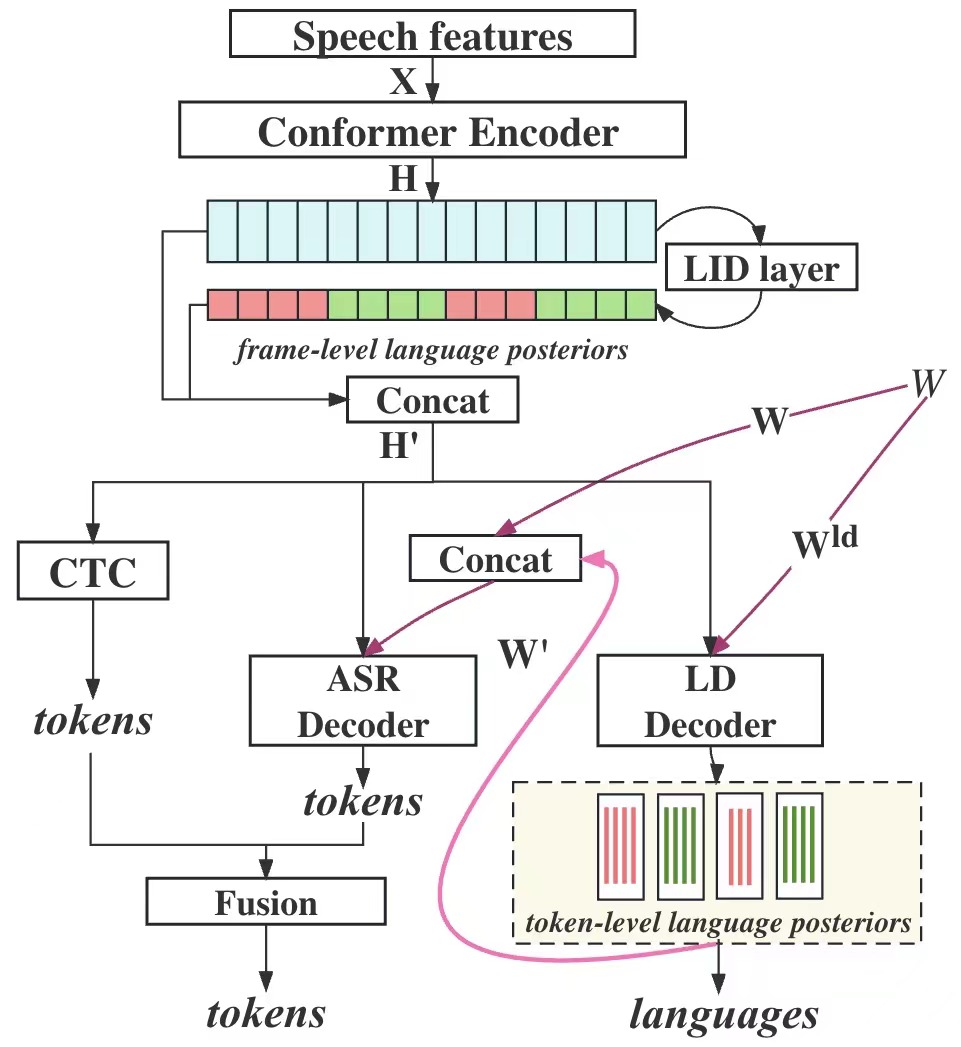}
  \caption{The hybrid CTC/attention model with interactive language biases.}
  \label{fig:joint_asr_ld}
\end{figure}

With reference to Fig.~\ref{fig:joint_asr_ld}, the frame-level language bias is achieved through a LID layer before being concatenated with the hidden output. In particular, the biased hidden output $\mathbf{H^{\prime}}$ is computed via
\begin{equation}
  \mathbf{h}^{\prime}_{t} = \mathrm{Concat}\big (\mathbf{h}_{t}, \mathbf{p}\left(l_{t}|\mathbf{h_{t}}\right) \big ),
  \label{eq:encoder_bias}
\end{equation}
which subsequently replaces $\mathbf{H}$ in (\ref{eq:asr_decoder}) to facilitate the interaction between language information among frames and tokens. In addition, $\mathbf{H^{\prime}}$ is also employed to develop a language-aware CTC module. The ASR decoder output is next achieved via
\begin{equation}
  p\left(w_{n}|w_{1:n-1},\mathbf{X}\right) = \mathrm{Decoder}\left ( \mathbf{w}^{\prime}_{1:n-1}, \mathbf{H^{\prime}} \right ).
  \label{eq:asr_decoder2}
\end{equation}

During training, the frame-level LID is optimized in an unsupervised manner similar to \cite{bi_encoder} (i.e., frame-level language annotations are not provided during training). However, frames in $\mathbf{H^{\prime}}$ are trained to be aligned with their corresponding token-level language labels within the language diarization decoder. An assumption made here is that an accurate frame-to-token alignment enriches the unsupervised LID process with supervised information through backpropagation. Optimization of the model is achieved similarly to that of (\ref{eq:loss_joint}). 

During inference, the frame- and token-level language posterior are computed before biasing the hidden output and the ASR decoding, respectively. The decoding process is similar to that presented in~\cite{hybrid_ctc_attention_asr}, which is defined to maximize the linear combination of the logarithmic CTC and attention objectives, i.e.,
\begin{equation}
\small
\setlength{\abovedisplayskip}{4pt}
\setlength{\belowdisplayskip}{4pt}
  \widehat{W} = \argmax_W \big \{ \alpha \mathrm{log} p_{\mathrm{ctc}}\left(W|\mathbf{X}\right) + \left ( 1-\alpha  \right ) \mathrm{log} p_{\mathrm{att}}\left(W|\mathbf{X}\right)  \big \}.
  \label{eq:decoding}
\end{equation}

\vspace{-0.2cm}
\section{Dataset, experiments, and results}
\label{sec:experiment}
\vspace{-0.2cm}
\subsection{Dataset and experiment setup}
All experiments are conducted on datasets from the ASRU 2019 Mandarin-English code-switching speech recognition
challenge \cite{shi2020asru}. This challenge comprises four datasets, including a 500-hour Mandarin-only training set, a 200-hour intra-sentence English-Mandarin code-switching training set, a 40-hour intra-sentence English-Mandarin code-switching development set, and a 20-hour intra-sentence English-Mandarin code-switching test set. We employed ESPnet~\footnote{Source code: https://github.com/Lhx94As/interactive\_language\_biases} to train all models on the 200-hour CS training set, which are validated on the development set and evaluated on the test set~\cite{espnet}.

SpecAugment is applied to augment the training data \cite{specaug}. Words are transformed into a total of $V=6,923$ tokens that include 3,000 English byte-pair encoding~(BPE) tokens, 3,920 Mandarin characters, and three special tokens for {\em unk}, {\em blank}, and {\em sos/eos}. All tokens are transformed to language labels building $\mathcal{V}^{\mathrm{ld}}$, which comprises {\em e} for English BPEs, {\em m} for Mandarin characters, and {\em sos/eos}. Language labels in $\mathcal{V}^{\mathrm{ld}}$ are used as LD outputs. We extracted $F=83$ dimensional features comprising 80-dimensional log-fbanks and 3-dimensional pitch for each speech sample before applying global mean and variance normalization.

We chose a hybrid CTC/Attention ASR model comprising twelve conformer encoder layers and six transformer decoder layers as the baseline model~\cite{conformer, transformer, karita19_interspeech}. In addition, we adopted the multi-task learning model and the language posterior bias approach as our benchmark~\cite{reduce_liu}. All self-attention encoder and decoder layers have four attention heads with input and output dimensions being $D=256$, and the inner layer of the position-wise feed-forward network is of 2048 dimensions. During training, we set parameters $\alpha=0.3$ and $\beta=0.8$ in (\ref{eq:loss_joint}), while a label smoothing factor of 0.1 is used for all cross-entropy losses. The ten best models during validation are averaged for inference. All models are trained on two GeForce RTX 3090 GPUs, where the baseline was trained for seventy epochs, while other models were trained for eighty epochs due to their higher number of parameters. 

During inference, we set parameters $\alpha=0.4$ in (\ref{eq:decoding}). Ten-best beam search is used before selecting the best hypothesis. The language model~(LM) used in this paper is a sixteen-layer transformer model with each attention layer comprising eight heads. The proposed systems are evaluated by employing mix error rate~(MER) comprising word error rate~(WER) for English and character error rate~(CER) for Mandarin.
\begin{table}[t]
\centering
\caption{Performance comparison of models utilizing various-level language information without using external language model by employing MER~(\%)}
\label{tab:language_bias}
\setlength{\tabcolsep}{0.95mm}{
\begin{tabular}{c|cc|c}
\hline
\textbf{Index} & \multicolumn{2}{c|}{\textbf{Method} }    & \textbf{MER}   \\ \hline
1.0     & \multicolumn{1}{c|}{Baseline}                     & Hybrid CTC/attention      & 12.8 \\ \hline
1.1   & \multicolumn{1}{c|}{Multi-task}                   & Multi-task with LD       & 12.4 \\ \hline
1.2   & \multicolumn{1}{c|}{Token-level}                  & Decoder LPB           & 12.4 \\ \hline
1.3   & \multicolumn{1}{c|}{Frame-level}                  & Encoder LPB               & 12.8 \\ \hline
1.4   & \multicolumn{1}{c|}{\multirow{3}{*}{Interactive}} & Encoder + CTC LPB         & 12.4 \\ \cline{1-1} \cline{3-4} 
1.5   & \multicolumn{1}{c|}{}                             & Encoder + Decoder LPB     & 12.1 \\ \cline{1-1} \cline{3-4} 
1.6   & \multicolumn{1}{c|}{}                             & Encoder+ Decoder + CTC LPB & \textbf{11.8} \\ \hline 
\end{tabular}}
\end{table}

\vspace{-0.2cm}
\subsection{Baseline and single language-biased models}
The results of the benchmark models are shown in Table~\ref{tab:language_bias} as systems 1.0, 1.1, and 1.2. Compared to the vanilla hybrid CTC-attention CS-ASR model, incorporating an auxiliary language diarization task and employing token-level LPB proposed in~\cite{reduce_liu} lead to higher performance. These indicate that incorporating language information benefits the CS-ASR process, which is consistent with the observation presented in~\cite{reduce_liu}. However, the token-level LPB approach shows no performance improvement over the multi-task optimization since Mandarin is the primary language in this dataset and languages do not switch frequently. 

The above data characteristics also result in performance degradation for model configuration 1.3 when frame-level LID is not sufficiently accurate. To prevent the CTC outputs from interacting with the unsupervised frame-level LID, the input of CTC is set to $\mathbf{H}$ while the input of the ASR decoder is set to $\mathbf{H}^{\prime}$ in model configuration 1.3. As mentioned in Section~\ref{sec:ilb}, frame-level LID is generally less accurate than token-level LID. Those incorrect language posteriors may increase language confusion when being transmitted into the ASR decoder module. 
\vspace{-0.2cm}
\subsection{Results of models with interactive language biases}
We next investigate how the interaction between frame- and token-level language information improves the model performance using systems 1.4, 1.5, and 1.6. In model configuration 1.4, as opposed to model configuration 1.3, the input of CTC is set to $\mathbf{H}^{\prime}$ so as to bias the CTC module with language information. The CTC performs frame-level classification before computing the optimal alignment, where the language biases are infused with acoustic features and combined intrinsically when generating tokens. Therefore, the performance improvement shown in Table~\ref{tab:language_bias} when comparing model configuration 1.4 with 1.3 indicates that language-biased frames can also perform better than vanilla frames. This underpins the efficacy of the frame-level language bias when being used for CTC. 

Model configuration 1.5 employs frame- and token-level language biases jointly but excludes the CTC module from being biased. Model configuration 1.5 shows significantly higher performance than single-language-biased models 1.2 and 1.3. This implies that the use of token-level language bias compensates for the inaccurate frame-level LID especially when model configuration 1.3 degrades the performance of model 1.1, which demonstrates that the interactive language biases are effective for CS-ASR. 

Model configuration 1.6 further combines two language biases with the CTC module and achieves the highest performance among all model considerations, with a 7.8\% relative improvement compared to the baseline model. It is not surprising that this configuration achieves higher performance than model configurations 1.4 and 1.5 since biasing CTC with language information improves the performance over the encoder LPB approach. In addition, the above implies that enriching all modules within a CS-ASR model with language information obtains higher gain compared to a single module, which is consistent with our proposition in Section~\ref{sec:intro}.
\begin{table}[t]
\centering
\caption{Performance comparison of models using external language model during inference by employing MER~(\%), where "Reduction'' denotes the absolute MER reduction compared to their no-LM counterparts}
\label{tab:language_modeling}
\setlength{\tabcolsep}{1mm}{
\begin{tabular}{c|c|c|c}
\hline
\textbf{Index} & \textbf{Method}   & \textbf{MER}  & \textbf{Reduction} \\ \hline
2.0   & Hybrid CTC/attention      & 12.6 & 0.2       \\ \hline
2.1   & Multi-task with LD        & 12.5 & -0.1      \\ \hline
2.2   & Decoder LPB               & 12.6 & -0.2      \\ \hline
2.4   & Encoder LPB               & 12.9 & -0.1      \\ \hline
2.5   & Encoder + CTC LPB         & 12.5 & -0.1      \\ \hline
2.6   & Encoder + Decoder LPB     & 12.3 & -0.2      \\ \hline
2.7   & Encoder + Decoder + CTC LPB & \textbf{11.9} & -0.1      \\ \hline
\end{tabular}}
\end{table}
\vspace{-0.2cm}
\subsection{Results of external language modeling}
Since the end-to-end ASR approaches internally perform language modeling, we explore whether the internal LM is stronger than the external LM when being trained on the same corpora. 

We present the results with respect to external language models in Table~\ref{tab:language_modeling}. The vanilla hybrid CTC/attention model shows higher performance after being integrated with external LM during inference. However, the results show that all language-aware CS-ASR models suffer from performance degradation compared to the baseline model. This implies that the CS-ASR model biased by language information could develop a more robust internal language model compared to an external model trained on the same text data. Since training an external language model can be time-consuming, robust internal language modeling can thus be concluded as an advantage of the proposed interactive language biases approach. 

\label{sec:method}
\begin{figure}[t]
\setlength{\belowcaptionskip}{-1cm}
  \centering
  \includegraphics[width=\linewidth]{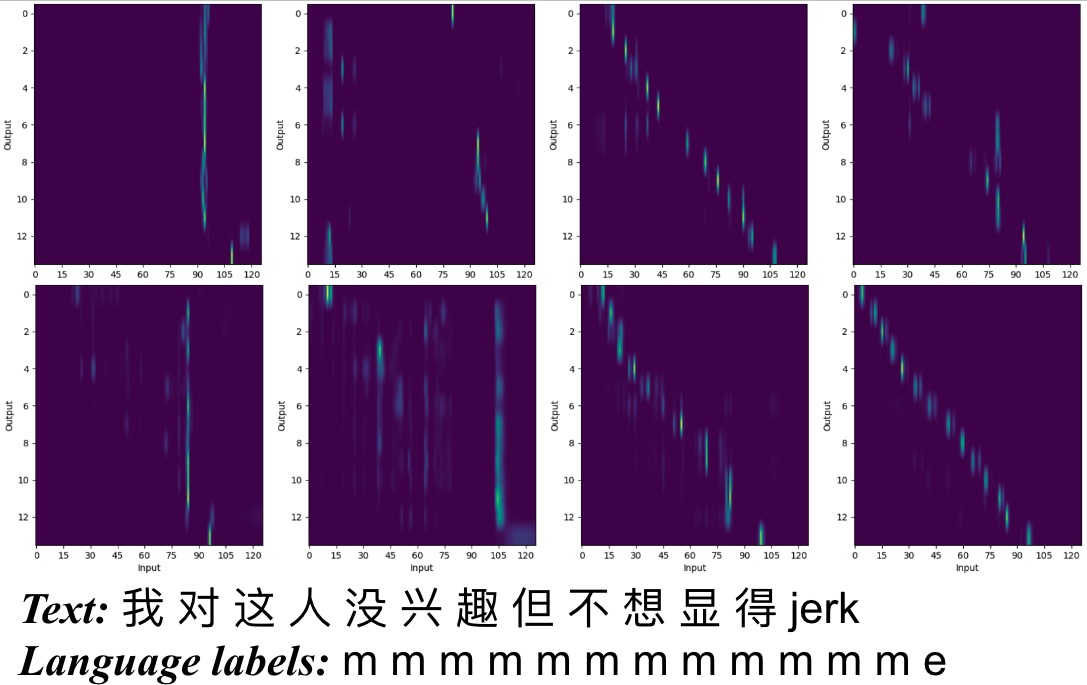}
  \caption{Comparison between attention matrices with respect to the frame-to-token alignment within language diarization decoder after employing token-level LPB (above) and interactive language biases (below).}
  \label{fig:comparison}
\end{figure}
\section{Discussion}
\label{sec:discussion}

Although the language diarization decoder adopted in this work does not generate timestamps for language changes, the frame-to-language alignment can be obtained from the attention matrices within the LD decoder as shown in Fig.~\ref{fig:comparison}.

The token-level LPB and interactive language biases (model configurations 1.2 and 1.6) are selected to compare single language bias with interactive language biases. As illustrated in Fig.~\ref{fig:comparison}, the attention mechanism identifies language changes in the first and second heads, and captures sequential information in the third and fourth heads. Compared to the token-level LPB, the attention matrices of our proposed interactive language biases approach exhibit clearer vertical language boundaries and smoother diagonal frame-to-token alignment. This indicates that the proposed approach improves not only ASR but also language diarization performance being consistent with our assumption in Section~\ref{sec:ilb}.
\section{Conclusion}
\label{sec:conclude}
We proposed an interactive language biases approach to improve CS-ASR through the interaction between frame- and token-level language information. Experiment results presented indicate that the proposed approach outperforms the benchmark in CS-ASR. We next visualized the attention matrices within the LD decoder. The proposed interactive language biases achieve higher language diarization performance compared with single token-level language bias, highlighting the efficacy of the proposed interactive language biases approach. In addition, the results show that a language-aware CS-ASR model can develop a robust internal LM, resulting in performance degradation when using an external language model during inference.

\vfill\pagebreak
\small
\bibliographystyle{IEEEtran}
\bibliography{strings, refs}

\end{document}